\documentclass[conference]{IEEEtran}
\IEEEoverridecommandlockouts
\usepackage{amsmath,amssymb,amsfonts}
\usepackage[utf8]{inputenc}
\usepackage{algorithm}
\usepackage[noend]{algpseudocode}
\algrenewcommand\algorithmicrequire{\textbf{Input:}}
\algrenewcommand\algorithmicensure{\textbf{Output:}}
\usepackage{array}
\usepackage{textcomp}
\usepackage{url}
\usepackage{graphics}
\usepackage{graphicx}
\usepackage{verbatim}
\usepackage{cite} 
\usepackage{siunitx}
\usepackage{booktabs, makecell}
\ifCLASSOPTIONcompsoc
    \usepackage[caption=false, font=normalsize, labelfont=sf, textfont=sf]{subfig}
\else
    \usepackage[caption=false, font=footnotesize]{subfig}
\fi
\usepackage{nomencl}
\usepackage{multirow,tabularx}
\usepackage{etoolbox}
\usepackage{array,multirow}
\usepackage{colortbl}
\usepackage{mathtools}
\usepackage{mdwtab}
\makeatletter
\newcommand*{\rom}[1]{\expandafter\@slowromancap\romannumeral #1@}
\makeatother
\usepackage{float}
\graphicspath{{./Figures/}}
\usepackage{comment}
\usepackage{amsmath,amssymb}
\usepackage{graphicx}
\usepackage{gensymb} 
\usepackage{bm} 
\usepackage{cleveref} 
\usepackage{booktabs}
\usepackage[table]{xcolor}
\makeatletter
\newcommand{\linebreakand}{%
  \end{@IEEEauthorhalign}
  \hfill\mbox{}\par
  \mbox{}\hfill\begin{@IEEEauthorhalign}
}
\makeatother

\begin{document}

\title{{Profile Screening and Frequency-Secure Operating Limit Derivation for Data Center Microgrids}\\
\thanks{This research is sponsored in part by M.C. Dean.}
}

\author{\IEEEauthorblockN{Zishuo Yang, Yu Ma, Jong Ha Woo, Aaditya Pandey,  \\ Wenyuan Tang, Chau-Wai Wong, Ning Lu}
\IEEEauthorblockA{
\textit{North Carolina State University}\\
Raleigh, NC 27606, USA \\
\{zyang66, nlu2\}@ncsu.edu}
\and
\IEEEauthorblockN{Andrew Meyers, George Navarro}
\IEEEauthorblockA{\textit{M.C. Dean} \\
Tysons, VA 22102, USA \\
\{andrew.meyers, george.navarro\}@mcdean.com}
}

\maketitle
\begin{abstract}
Rapid power variations from artificial intelligence (AI) workloads pose frequency-security challenges for islanded data-center microgrids. This paper proposes a generator step-response-based framework to rapidly screen candidate load profiles and derive frequency-secure operating limits for IT workload scheduling. Identified generator responses are superimposed to assess the cumulative frequency effects of successive load changes without repeated dynamic simulations. The framework identifies potentially unsafe intervals and determines allowable load magnitudes, ramp rates, and minimum event spacing, while accounting for prediction uncertainty and governor limits. Validation on a 2.5~MVA islanded microgrid with eight generator configurations and 14 synthetic AI workload profiles shows that 94.75\% of 40,208 assessment windows receive decisive screening labels, with no frequency-violating windows classified as secure.
\end{abstract}


\color{black}

\section{Introduction}

The rapid growth of artificial intelligence (AI) data centers
is introducing large and highly dynamic electrical loads
\cite{nerc2025largeloads}. Grid capacity constraints and
long infrastructure development timelines have increased
interest in on-site generation and microgrid-based power
supply \cite{olivares2014trends}. However, during islanded
operation, local generators have limited capability to
accommodate rapid load variations due to their inertia,
governor dynamics, available power headroom, and actuator
constraints. Consequently, rapid and repetitive power
swings from AI workloads can cause excessive frequency
excursions and rate-of-change-of-frequency (RoCoF)
violations \cite{ko2026aioscillations}.

Workload scheduling and GPU power control can adjust data-center demand \cite{colangelo2026grid,zhou2026grid}. However, in an islanded microgrid, scheduling a workload pause and its subsequent resumption requires knowing both how much power variation the generators can safely accommodate and how candidate load profiles affect frequency security. These requirements present two
related challenges: rapidly identifying potentially unsafe intervals in dynamic load profiles and translating generator capabilities into allowable load magnitudes, ramp rates, and event spacing. Closely spaced load transitions can produce overlapping frequency responses, making individual step-size and ramp-rate limits insufficient to assess their cumulative effects. Waveform-level screening can provide timely feedback for workload scheduling, enabling targeted adjustments to task timing and power demand before execution.

\begin{figure*}[t]
  \centering
  \includegraphics[width=0.9\textwidth]{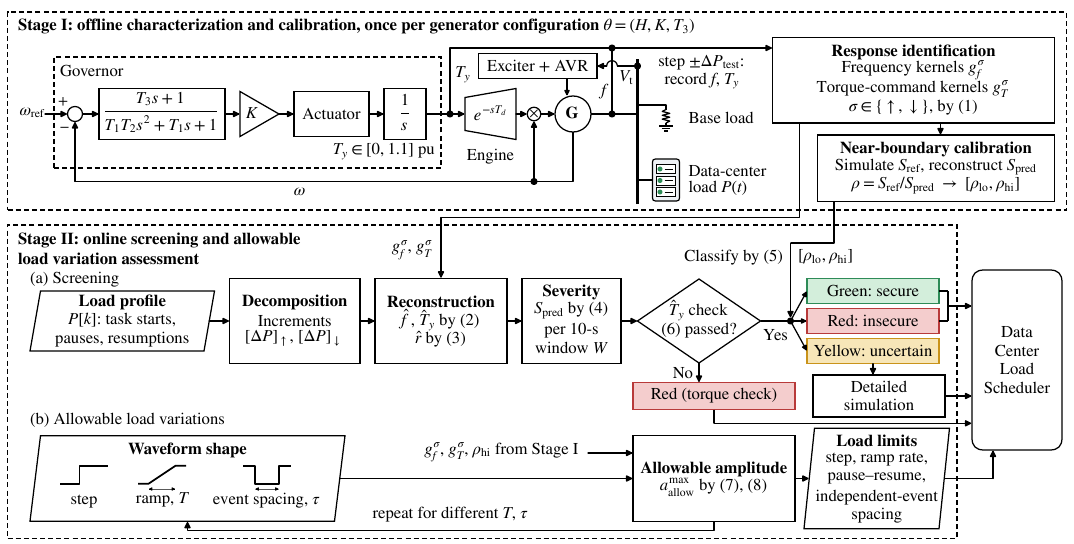}
  \setlength{\abovecaptionskip}{-1pt}
\caption{Flowchart of the proposed two-stage framework for rapid frequency-security screening and allowable load variation assessment in data-center microgrids. The online stage (a) screens candidate load profiles and (b) derives allowable load variations.}
  \label{fig:overview}
  \vspace{-6pt}
\end{figure*}

Detailed dynamic simulations can assess the frequency response to a specified load profile
\cite{pnnl_large_load_oscillation,hyperscale_dc_stability}, but finding the allowable amplitude may require several runs even when the waveform shape and event timing are fixed. Reduced-order analytical methods assess frequency response following individual disturbances~\cite{sfr_analytical,nadir_prediction}, and data-driven methods have been used to incorporate post-outage frequency constraints into unit commitment \cite{ml_frequency_nadir}. However, data-center workload scheduling requires assessing the combined frequency response to successive load changes and deriving the maximum allowable load amplitudes for each waveform and timing.

This paper addresses both tasks by developing a rapid waveform-level frequency-security screening method that identifies potentially unsafe intervals in data-center load profiles by capturing the cumulative effects of successive load changes through generator step-response superposition. Empirical error calibration and governor torque-command checks support
screening decisions, with uncertain cases labeled for detailed dynamic simulation. More importantly, the framework translates generator frequency-security capabilities into allowable load-profile operating limits. For each prescribed waveform shape and event timing, it derives the allowable load amplitude by scaling the predicted response at a single reference amplitude. Repeating this calculation over different ramp durations and event spacings yields allowable step magnitudes and ramp rates, a pause–resume limit for two opposite load changes, such as a workload pause followed by its resumption, and an independent-event spacing beyond which successive changes can be treated as independent.

\section{Methodology}\label{sec:method}

Fig.~\ref{fig:overview} illustrates the proposed two-stage framework. In the offline stage, load-increase and load-decrease tests in MATLAB/Simulink identify the generator’s frequency and governor torque-command step responses. Comparisons between simulated and predicted severity near the frequency-security boundary establish empirical bounds on prediction error for each generator configuration.

In the online stage, each candidate load profile is decomposed into signed power increments, and the corresponding step responses are superimposed to reconstruct the frequency and torque-command trajectories. Branch (a) screens each assessment window using thresholds derived from the empirical error bounds and a governor torque-command check. Windows that fail the torque-command check are labeled red, and uncertain windows are referred to detailed simulation. Branch (b) uses the same step responses to derive allowable step magnitudes, ramp rates, pause–resume limits, and independent-event spacing for data-center load scheduling.

\subsection{Generator Response Characterization}
For a given generator configuration $\boldsymbol{\theta}$,
the generator's dynamic characteristics are identified
through load-increase and load-decrease step tests using the detailed Simulink model. The generator is initialized at
a steady operating point $(P_0,f_0,T_{y0})$, where $T_y$
denotes the governor torque command before the engine delay.
Both frequency and torque-command responses are recorded,
with step magnitudes selected to avoid governor saturation.

Let $\mathbf{y}=[f,T_y]^{\mathsf{T}}$. The normalized
step-response kernels are defined as
\begin{equation}
\mathbf{g}^{\sigma}(t;\boldsymbol{\theta})
=
\frac{\mathbf{y}^{\sigma}(t)-\mathbf{y}_0}
{\Delta P_{\mathrm{test}}^{\sigma}},
\quad \sigma\in\{\uparrow,\downarrow\},
\label{eq:kernel}
\end{equation}
where $\mathbf{y}^{\sigma}(t)$ is the response recorded in the load-increase ($\sigma={\uparrow}$) or load-decrease ($\sigma={\downarrow}$) test, $\mathbf{y}_0=[f_0,T_{y0}]^{\mathsf{T}}$ is the initial steady state, and the test step $\Delta P_{\mathrm{test}}^{\sigma}$ is positive for load increases and negative for load decreases. The kernel $\mathbf{g}^{\sigma}=[g_f^{\sigma},g_T^{\sigma}]^{\mathsf{T}}$ contains the frequency and torque-command components and is stored for screening.

\begin{figure}[!t]
\centering
\includegraphics[width=\columnwidth]{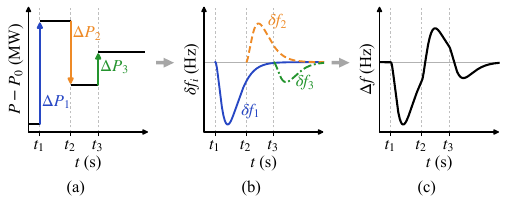}
\setlength{\abovecaptionskip}{-15pt}
\caption{Illustration of response reconstruction:
(a) decomposition of a load profile into signed steps;
(b) scaling and shifting of individual step responses;
and (c) reconstruction of the frequency response through superposition.}
\label{fig:superposition}

\end{figure}

\subsection{Duhamel-Based Response Reconstruction}
An arbitrary load profile is decomposed into positive and negative load steps. Assuming approximately linear generator dynamics, the frequency and torque-command responses are reconstructed by scaling, shifting, and superimposing the identified step-response kernels according to each load change’s magnitude, direction, and timing, as illustrated in Fig.~\ref{fig:superposition}. Its accuracy is evaluated in Section~\ref{subsec:verification}.

For a sampled load profile, the load increments are
$\Delta P[k]=P[k]-P[k-1]$. The reconstructed response
is expressed using the discrete form of Duhamel's principle:
\begin{equation}
\hat{\mathbf{y}}[n]=\mathbf{y}_0+\sum_{\sigma\in\{\uparrow,\downarrow\}}\sum_{k=1}^{n}\mathbf{g}^{\sigma}[n-k]\,[\Delta P[k]]_{\sigma},
\label{eq:recon}
\end{equation}
where $\mathbf{g}^{\sigma}[m]$ denotes the sampled step-response kernel, and $[x]_{\uparrow}=\max(x,0)$ and $[x]_{\downarrow}=\min(x,0)$ denote positive and negative load increments, respectively.

The reconstruction captures cumulative effects of
closely spaced load transitions, including ramps
represented by successive steps, while preserving
the steady-state torque-command changes associated
with sustained load variations.

\subsection{Frequency Assessment and Torque-Command Check}
\label{sec:assess}

The reconstructed frequency trajectory is assessed against frequency nadir, peak, and RoCoF limits. RoCoF is estimated as the least-squares slope over a moving 0.5~s window:
\begin{equation}
\hat{r}[n]=\frac{\sum_{j=-m}^{m}(j\Delta t)\,\hat{f}[n+j]}{\sum_{j=-m}^{m}(j\Delta t)^{2}},
\label{eq:rocof}
\end{equation}
where $\Delta t$ is the sampling interval and $2m\Delta t=0.5$~s. The same estimator is applied to the reconstructed and simulated frequency trajectories.

For each assessment window $\mathcal{W}$, the predicted frequency-security severity is defined as
\begin{equation}
S_{\mathrm{pred}}=\max\left\{\frac{f_{\mathrm{n}}-\hat{f}_{\mathrm{min}}}{f_{\mathrm{n}}-f_{\mathrm{L}}},\,
\frac{\hat{f}_{\mathrm{max}}-f_{\mathrm{n}}}{f_{\mathrm{U}}-f_{\mathrm{n}}},\,
\frac{\hat{R}}{r_{\mathrm{lim}}}\right\},
\label{eq:severity}
\end{equation}
where $\hat{f}_{\mathrm{min}}$, $\hat{f}_{\mathrm{max}}$, and $\hat{R}$ are the predicted frequency minimum, frequency maximum, and maximum absolute RoCoF within $\mathcal{W}$, respectively. Here, $f_{\mathrm{n}}$ is the nominal frequency, $f_{\mathrm{L}}$ and $f_{\mathrm{U}}$ are the lower and upper frequency limits, and $r_{\mathrm{lim}}$ is the RoCoF limit. A severity of 1 marks the frequency-security boundary.

Near this boundary, even a small prediction error can change the screening decision. To quantify this error, the same severity index is evaluated using the frequency trajectory from the detailed Simulink model to obtain the reference severity $S_{\mathrm{ref}}$. For each generator configuration, the minimum and maximum values of the ratio $\rho=S_{\mathrm{ref}}/S_{\mathrm{pred}}$ observed in cases near the boundary define the empirical severity-ratio interval $[\rho_{\mathrm{lo}},\rho_{\mathrm{hi}}]$. Their coverage on new load profiles is evaluated in Section~\ref{subsec:profile_validation}.

For each assessment window, multiplying the ratio bounds by $S_{\mathrm{pred}}$ gives the empirical severity interval $[\rho_{\mathrm{lo}}S_{\mathrm{pred}},\,\rho_{\mathrm{hi}}S_{\mathrm{pred}}]$. If the window passes the torque-command check described below, it is labeled green when the entire interval lies below 1, red when the entire interval lies at or above 1, and yellow otherwise, in which case detailed simulation is needed. In terms of $S_{\mathrm{pred}}$, these rules are
\begin{equation}
\begin{cases}
\text{Green}, & S_{\mathrm{pred}}<1/\rho_{\mathrm{hi}},\\
\text{Yellow}, & 1/\rho_{\mathrm{hi}}\le S_{\mathrm{pred}}<1/\rho_{\mathrm{lo}},\\
\text{Red}, & S_{\mathrm{pred}}\ge 1/\rho_{\mathrm{lo}}.
\end{cases}
\label{eq:labels}
\end{equation}

Response reconstruction is valid only while the governor torque command remains within its limits. A window passes this check only if
\begin{equation}
\underline{T}_{y}+\epsilon<\hat{T}_{y}[n]<\overline{T}_{y}-\epsilon,\qquad \forall\, t_n\in\mathcal{W},
\label{eq:torque}
\end{equation}
where $\underline{T}_{y}$ and $\overline{T}_{y}$ are the lower and upper governor torque-command limits and $\epsilon$ is the margin from these limits. A window that fails this check is labeled red. Once the torque command reaches its limit, the governor can no longer increase or decrease the mechanical power and loses its ability to correct the frequency deviation, and the superposition of step responses no longer holds.

The frequency-security and governor torque-command constraints are used next to derive allowable load variations.

\subsection{Derivation of Allowable Load Variations}

For a given generator configuration and operating point,
the step-response kernels are used to determine the maximum allowable amplitude of a load waveform with fixed shape and timing, starting from a steady operating point. The amplitude $a$ denotes the size of each load change in the waveform. Under approximately linear dynamics and nominal initial frequency, the predicted severity scales with $a$:
\begin{equation}
S_{\mathrm{pred}}(a)
=
\frac{a}{a_{\mathrm{ref}}}
S_{\mathrm{pred}}(a_{\mathrm{ref}}),
\label{eq:severity_scaling}
\end{equation}
where $a_{\mathrm{ref}}$ is the reference amplitude and
$a$ scales all load deviations proportionally.

A window remains green under~\eqref{eq:labels} only if $\rho_{\mathrm{hi}} S_{\mathrm{pred}}(a) < 1$. Combining this condition with the torque-command check~\eqref{eq:torque} gives the maximum allowable amplitude
\begin{equation}
a_{\mathrm{allow}}^{\max}
=
\min\left\{
\frac{a_{\mathrm{ref}}}
{\rho_{\mathrm{hi}}S_{\mathrm{pred}}(a_{\mathrm{ref}})},
a_T
\right\},
\label{eq:allowable_amplitude}
\end{equation}
where $a_T$ is the largest amplitude for which the reconstructed torque command satisfies~\eqref{eq:torque} over the entire trajectory.

\begin{table*}[!t]
\centering
\caption{Testbed Configurations, Frequency-Security Screening
Results, and Allowable Load Variations}
\label{tab:combined_results}
\renewcommand{\arraystretch}{1.15}
\setlength{\tabcolsep}{3pt}
\begin{tabular*}{\textwidth}{
@{\extracolsep{\fill}}cccccccccccccccc@{}}
\toprule
& \multicolumn{3}{c}{\textbf{Generator}}
& \multicolumn{2}{c}{\textbf{Calibration}}
& \multicolumn{6}{c}{\textbf{Frequency-Security Screening}}
& \multicolumn{4}{c}{\textbf{Allowable Load Variations}} \\
\cmidrule(lr){2-4}
\cmidrule(lr){5-6}
\cmidrule(lr){7-12}
\cmidrule(l){13-16}
\textbf{Case}
& $H$ & $K$ & $T_3$ & $\rho_{\rm lo}$ & $\rho_{\rm hi}$
& Green & Yellow & Red & Missed & False & Green + Red
& Step & Ramp rate & Pause--resume & Independent \\
& (s) & & (s)
& & 
& & &
& violations & alarms & (\%)
& (kW) & (kW/s) & (kW) & spacing (s) \\
\midrule
A & 1.00 & 20 & 0.40 & 0.9437 & 1.0290
  & 1773 & 323 & 2930 & 0 & 0 & 93.57
  & 276 & 402 & 196 & 1.7 \\
B & 1.00 & 14 & 0.40 & 0.9513 & 1.0593
  & 1403 & 67 & 3556 & 0 & 0 & 98.67
  & 209 & 275 & 146 & 1.9 \\
C & 1.00 & 20 & 0.60 & 0.9344 & 1.0712
  & 2229 & 549 & 2248 & 0 & 12 & 89.08
  & 332 & 399 & 248 & 1.9 \\
D & 1.50 & 20 & 0.40 & 0.9195 & 1.0461
  & 1989 & 294 & 2743 & 0 & 0 & 94.15
  & 310 & 383 & 216 & 1.9 \\
E & 1.50 & 14 & 0.40 & 0.9221 & 1.0572
  & 1655 & 90 & 3281 & 0 & 0 & 98.21
  & 250 & 263 & 172 & 2.1 \\
F & 1.50 & 20 & 0.60 & 0.9109 & 1.0485
  & 2590 & 475 & 1961 & 0 & 0 & 90.55
  & 381 & 402 & 282 & 2.1 \\
G & 0.75 & 14 & 0.40 & 0.9668 & 1.0563
  & 1177 & 224 & 3625 & 0 & 0 & 95.54
  & 192 & 283 & 135 & 1.7 \\
H & 0.75 & 20 & 0.40 & 0.9576 & 1.0263
  & 1736 & 89 & 3201 & 0 & 0 & 98.23
  & 257 & 410 & 184 & 1.6 \\
\midrule
\textbf{Total}
& -- & -- & -- & -- & --
& 14552 & 2111 & 23545 & 0 & 12 & 94.75
& -- & -- & -- & -- \\
\bottomrule
\end{tabular*}
\par\vspace{2pt}
\begin{minipage}{\textwidth}
\footnotesize\raggedright
\end{minipage}
\vspace{-18pt}
\end{table*}

Equation~\eqref{eq:allowable_amplitude} is applied to the load variations used in workload scheduling. The step limit is the allowable amplitude of a single load change and bounds an abrupt power change at workload start or stop. For a linear ramp of duration $T$, dividing the frequency-limited amplitude $a_\mathrm{ref}/(\rho_\mathrm{hi} S_\mathrm{pred}(a_\mathrm{ref}))$ by $T$ gives the ramp-rate limit. This rate decreases as $T$ increases and is therefore evaluated at the longest duration considered, while $a_T$ bounds the total load increase. For two successive changes,~\eqref{eq:allowable_amplitude} gives the allowable amplitude per event as a function of their separation $\tau$. If the two changes are opposite, as in a workload pause followed by its resumption, the pause–resume limit is the smallest such amplitude over all evaluated $\tau$ and both orders. This worst case keeps the frequency secure regardless of the pause duration. The independent-event spacing is the smallest $\tau$ beyond which the allowable amplitude remains at least a prescribed fraction of the single-step limit.

\section{Simulation Results}

\label{sec:case_study}
A 2.5~MVA islanded diesel microgrid is modeled
in MATLAB/Simulink with isochronous governor control
and an automatic voltage regulator (AVR).
The generator supplies a 0.1~MW constant-impedance
base load and a time-varying constant-power
data-center load. The governor torque command $T_y$
is the limited integrator output before the engine
delay. Frequency limits are set to 59.4--60.6~Hz,
with a RoCoF limit of 1~Hz/s. Simulations use a discrete phasor model with an 8.33~ms solver step, which also serves as the sampling interval, a 0.5~s RoCoF estimation window, and 10~s assessment windows. The torque-command limits are 0 to 1.1~pu, with a margin of $\epsilon=0.011$~pu.

Eight generator configurations (A--H), listed in Table~\ref{tab:combined_results}, are evaluated
by varying the inertia constant $H$, governor gain
$K$, and lead time constant $T_3$, while keeping
other parameters fixed. Configuration A
($H=1.0$~s, $K=20$, $T_3=0.4$~s) serves as
the baseline. For each configuration, $\pm150$~kW
load-step tests at $P_0=1$~MW identify the frequency
and torque-command responses without governor
saturation.

For each configuration, 144 calibration cases covering single, consecutive, and periodic load variations are generated at 1~MW and at the upper and lower edges of the load range (2.35 and 0.25~MW), all reconstructed with the kernels identified at 1~MW.
Load amplitudes are adjusted to obtain predicted
severity values of $0.85\leq S_{\mathrm{pred}}\leq1.15$,
providing near-boundary cases for estimating the
empirical severity-ratio bounds reported in Table~\ref{tab:combined_results}.

\subsection{Verification of Amplitude Scaling and Superposition}
\label{subsec:verification}

The homogeneity and superposition assumptions are evaluated using configuration G ($H = 0.75$~s, $K = 14$, $T_3 = 0.4$~s). The 18 amplitude-scaling cases apply load increases and decreases of 75, 300, and 450~kW as steps or as 0.5~s and 1~s ramps. Each simulated response is compared with the $\pm$150~kW identification response scaled to the same amplitude. The 24 consecutive load-change cases (12 same-direction and 12 opposite-direction) consist of two 150 or 300~kW changes separated by $\tau = 0.2$--$2.0$~s. 

\begin{figure}[t] 
    \centering
    \includegraphics[width=0.45\textwidth]{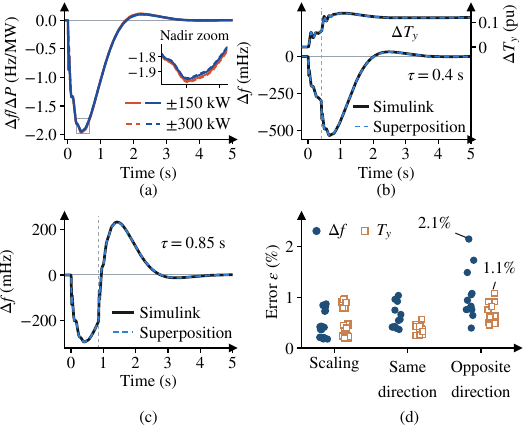}
    \setlength{\abovecaptionskip}{-1pt}
\caption{Verification of approximate generator-response
linearity: (a) amplitude scaling: normalized responses to the ±150 kW identification steps and $\pm$300~kW test steps; (b) superposition of same-direction load changes for $\Delta f$ and $\Delta T_y$;
(c) superposition of opposite-direction load changes; and
(d) maximum normalized trajectory errors of $\Delta f$ and $T_y$ across 42 test cases.}
    \label{fig:linearity}
    \vspace{-10 pt}
\end{figure}

Fig.~\ref{fig:linearity} compares the reconstructed and detailed simulated frequency responses in panels (a)--(c) and the torque command in panel (b).
The maximum trajectory error is normalized by the peak
simulated deviation of each response variable.
As summarized in Fig.~\ref{fig:linearity}(d), the maximum errors across all 42 cases are 2.1\% for the frequency deviation $\Delta f$ and 1.1\% for the torque command $T_y$. These results support the
approximate homogeneity and superposition assumptions
within the tested operating range.

\subsection{Validation of Load-Profile Screening}
\label{subsec:profile_validation}

The proposed method is evaluated using 14 synthetic
1~h load profiles representing pretraining, inference,
fine-tuning, and diffusion workloads. Eight profiles
represent individual workloads at two power scales,
while six represent pairwise workload combinations.
Each profile contains 359 assessment windows, yielding
5,026 windows per generator configuration and 40,208
windows across all eight configurations.

\begin{figure}[h] 
\centering
\includegraphics[width=\columnwidth]{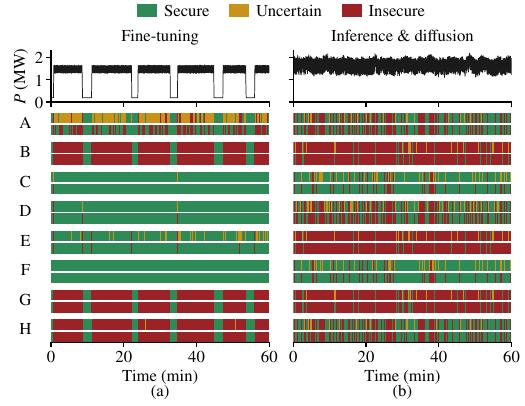}
\setlength{\abovecaptionskip}{-15pt}
\caption{Window-level frequency-security screening across eight generator configurations for (a) fine-tuning and (b) mixed inference and diffusion load profiles. For each configuration, the upper strip shows the predicted screening labels and the lower strip shows the Simulink frequency-security outcomes.}
\label{fig:profile_screening}

\end{figure}

Fig.~\ref{fig:profile_screening} compares the screening results with Simulink frequency-security outcomes for two representative profiles. For the fine-tuning profile, configurations B, G, and H violate the frequency-security limits during most high-load intervals, whereas configurations C, D, and F remain secure, showing that the same workload can be secure or insecure depending on generator dynamics. In this profile, configuration A operates close to the frequency-security boundary, so 60\% of its windows are labeled yellow and referred to detailed simulation. For the mixed inference and diffusion profile, violations alternate with secure windows within minutes, and the predicted labels follow these transitions.

Table~\ref{tab:combined_results} summarizes the window-level screening results. No window that violated the frequency-security limits in Simulink was labeled green. The 2,111 yellow windows contain both violating (1,214) and secure (897) cases, confirming the need for detailed simulation. 

Of the 286 windows that failed the torque-command check, 274 violated the frequency limits. The remaining 12, all under configuration C, were false alarms with predicted torque commands only slightly beyond the margin. Among the 5,218 near-boundary windows ($0.85 \le S_\mathrm{pred} \le 1.15$) that passed this check, only one severity ratio fell outside the calibrated interval, and its label remained correct. 

Overall, 94.75\% of the 40,208 assessment windows received a green or red screening label. With the precomputed step responses, screening a 1~h load profile took approximately 1.2~s. A full simulation of the same profile using the detailed Simulink model took approximately 136~s.

\subsection{Frequency-Secure Load Profile Variations}

For each prescribed waveform shape and event timing, \eqref{eq:allowable_amplitude} gives the allowable load amplitude by scaling the response reconstructed at a single reference amplitude. Table~\ref{tab:combined_results} reports the resulting load and timing limits for eight generator configurations, each evaluated from a steady operating point of $P_0 = 1$~MW. The step and pause--resume limits are 1.8\%--6.6\% below the corresponding boundary amplitudes in Simulink. The limits derived from a single reference response therefore remain close to the simulated boundaries, with a small conservative margin introduced by $\rho_\mathrm{hi}$ in \eqref{eq:allowable_amplitude}.

The allowable pause--resume amplitudes range from 135 to 282~kW, corresponding to 68.8\%--74.7\% of the respective single-step limits. This reduction arises because the two responses overlap: when the second change occurs, the governor is still responding to the first, which amplifies the resulting frequency deviation, so a change that is secure as an isolated step can become insecure when followed by an opposite change at certain spacings. The allowable single-step magnitudes range from 192 to 381~kW, and the ramp-rate limits range from 263 to 410~kW/s, depending on generator configuration. 

The independent-event spacings range from 1.6 to 2.1~s, beyond which the allowable amplitude of each change remains at least 90\% of the single-step limit and successive load changes can be treated as independent. The resulting amplitude and spacing limits can guide the scheduling of workload activation, suspension, and resumption.

\section{Conclusion}

This paper developed a generator step-response-based framework for rapid profile frequency-security screening and allowable load variation assessment in islanded data-center microgrids. By capturing the cumulative effects of successive load changes through response superposition, the framework identifies potentially unsafe workload intervals without repeated dynamic simulations. Across eight generator configurations and 14 synthetic AI workload profiles, 94.75\% of 40,208 assessment windows were classified as secure or insecure without detailed simulation, with no frequency-violating windows classified as secure. 

The framework also translates generator frequency-security capabilities into workload scheduling constraints by scaling a single reference response for each prescribed waveform and event timing. The constraints bound power steps at workload start or stop, ramp rates during gradual power adjustments, and power swings during pause--resume sequences. They also specify the minimum spacing for treating successive workload events as independent.

Our future work will extend the framework to multi-generator
and battery-supported microgrids, incorporate nonlinear
generator responses, and integrate screening with
automated workload scheduling.

\bibliographystyle{IEEEtran}
\bibliography{reference}

\end{document}